\documentclass{ws-ijgmmp}
\begin{document}
\markboth{E.~Elizalde, S.D.~Odintsov, E.O.~Pozdeeva, S.Yu.~Vernov}
{De Sitter and Power-law Solutions in Non-local Gauss-Bonnet Gravity}

\title{\textbf{De Sitter and Power-law Solutions in Non-local Gauss-Bonnet Gravity}}
\author{E.~Elizalde}
\address{1) Instituto de Ciencias del Espacio (ICE/CSIC) \\ Campus UAB, Carrer de Can Magrans, s/n
  08193 Bellaterra (Barcelona), Spain\\
 2) Institut d'Estudis Espacials de Catalunya (IEEC) \\
 Campus UAB, Carrer de Can Magrans, s/n
  08193 Bellaterra (Barcelona), Spain\\
  3) Laboratory for Theoretical Cosmology, Tomsk State University of
Control Systems and Radioelectronics (TUSUR), 634050 Tomsk, Russia \\
4) Tomsk State Pedagogical University, 634061 Tomsk, Russia
   \\ \email{elizalde@ieec.uab.es}}
 \author{S.D.~Odintsov}
\address{1) Instituci\'o Catalana de Recerca i Estudis Avancats
(ICREA), Barcelona, Spain \\
2) Instituto de Ciencias del Espacio (ICE/CSIC) \\ Campus UAB, Carrer de Can Magrans, s/n
  08193 Bellaterra (Barcelona), Spain\\
 3) Institut d'Estudis Espacials de Catalunya (IEEC) \\
 Campus UAB, Carrer de Can Magrans, s/n
  08193 Bellaterra (Barcelona), Spain \\
 4) Laboratory for Theoretical Cosmology, Tomsk State University of
Control Systems and Radioelectronics (TUSUR), 634050 Tomsk, Russia \\
5) Tomsk State Pedagogical University, 634061 Tomsk, Russia\\
 \email{odintsov@ieec.uab.es}}

\author{E.O.~Pozdeeva}
\address{Skobeltsyn Institute of Nuclear Physics, Lomonosov Moscow State University,\\ Leninskiye Gory~1, 119991, Moscow, Russia\\
\email{pozdeeva@www-hep.sinp.msu.ru}}

\author{S.Yu.~Vernov}
\address{{Skobeltsyn Institute of Nuclear Physics, Lomonosov Moscow State University,\\ Leninskiye Gory~1, 119991, Moscow, Russia}\\
\email{svernov@theory.sinp.msu.ru}}

\date{ \ }
\maketitle
\begin{abstract}
The cosmological dynamics of a non-locally corrected gravity theory, involving a
power of the inverse d'Alembertian, is investigated. Casting the dynamical
equations into local form, the fixed points of
the models are derived, as well as corresponding de Sitter and power-law solutions.
Necessary and sufficient conditions on the model parameters  for the existence of de Sitter solutions are obtained.
The possible existence of power-law solutions is investigated, and it is proven that models with de Sitter solutions
have no power-law solutions. A model is found, which allows to describe the matter-dominated phase of the Universe evolution.
\end{abstract}


\section{Introduction}
Reliable astronomical data support the existence of four distinct epochs of the Universe global evolution: inflation, a radiation dominated era, a matter dominated one and the present dark energy epoch. Initial inflation and dark energy domination are both characterized by an accelerated expansion of the Universe with almost constant Hubble parameter $H$. The other epochs of the Universe evolution are described by  power-law solutions with $H=J/t$, where $J$ is a positive constant. In General Relativity, power-law solutions with $H=J/t$ correspond to models with a perfect fluid whose EoS parameter reads $w_{\mathrm{m}}=-1+2/(3J)$. As a consequence, the radiation dominated epoch corresponds to solutions with $J=1/2$, whereas the matter dominated one corresponds to  solutions such that $J=2/3$. When one addresses the issue to consider new modified gravity models, it is therefore  important to check for the existence of de Sitter and power-law solutions in the discussed models.

There are two basic motivations which lead cosmologists to modify gravity. The first one is an attempt to connect gravity with quantum physics, at least in a perturbative way, by including quantum correction terms to Einstein's equations. The second is an interest to describe the Universe evolution in a more natural way, without the dark energy and the dark matter components, which turn out to be avoidable in the modified models.

In string theory inspired models that include extra dimensions the Gauss-Bonnet term arises naturally~\cite{Antoniadis:1993jc,Torii:1996yi,Kawai1998}.
Such types of models allow to describe different epochs of the cosmological evolution. In particular, Gauss-Bonnet gravity models are actively used to investigate either the dark energy domination era~\cite{Nojiri:2005jg,Cognola:2006eg,Cognola:2006sp,Sami:2005zc,Elizalde:2010jx,Benetti:2018zhv},  the inflation period~\cite{Guo:2010jr,Oikonomou:2017ppp}, or these two stages of the Universe evolution under the same scheme~\cite{Cognola:2009jx,vandeBruck:2017voa}. The scalar-Gauss-Bonnet gravity models allow to reconstruct  the Universe expansion history, in which the matter dominated era makes a transition towards the cosmic acceleration epoch~\cite{Cognola:2006sp}. The $f(R)$-Gauss-Bonnet gravity, described by an analytic function of the Ricci scalar $R$ and the Gauss-Bonnet invariant, is being actively investigated as well. In particular, the Noether symmetry approach that allows to find out exact solutions has been developed in~\cite{Capozziello:2014ioa},  the Newtonian and post Newtonian limits have been analyzed in~\cite{DeLaurentis:2013ska}. Note that both inflationary and dark energy evolutions can be described in the context of the $f(R)$-Gauss-Bonnet gravity~\cite{DeLaurentis:2015fea,Odintsov:2018nch}.

Another type of  intensely investigated, modified gravity models is non-local gravity~\cite{Non-local-gravity-Refs}. Most of the non-local cosmological models are inspired by string theory~\cite{Non-local_scalar} or by quantum field theory~\cite{Deser:2007jk,Tsamis:2014hra,Barvinsky:2014lja}.
Usually, non-local models include an analytic function of  either the d'Alembertian operator $\Box$ or the inverse d'Alembertian operator $\Box^{-1}$. Note that models including $F(\Box R,\Box^2 R,\dots,\Box^{-1}R,\Box^{-2}R,\dots)$ have been investigated as well~\cite{Jhingan:2008ym,Kluson}.

A model that includes a non-local
 term as a function of the $\Box^{-1}$ operator: $R f(\Box^{-1}R)$   has been proposed in~\cite{Deser:2007jk} to explain current cosmic acceleration without dark energy (see also~\cite{Woodard,Deser:2013uya,Woodard:2014iga}).
This class of non-local gravity models can be reformulated as a class of local model with additional scalar fields non-minimally coupled to gravity~\cite{Odintsov0708}. It has been shown that the non-local and local formulations give the same solutions for the linear perturbations as long as the initial conditions are set the same~\cite{Park:2017zls}.
The relation between the original non-local equations and their biscalar-tensor representation in the context of an acausality problem has been discussed in~\cite{Zhang:2016ykx}. The difference between these formulations has been emphasized in~\cite{Deser:2013uya}.
 The ensuing cosmology describing the four basic epochs and the perturbation analysis were studied in~\cite{Koivisto}. Structure formation in this non-locally modified gravity model is being actively discussed~\cite{Park:2017zls,ParkDonelson,Nersisyan:2017mgj}. The most studied model of this type is characterized by an exponential function $f(\Box^{-1}R)=f_0 e^{\alpha(\Box^{-1}\!R)}$, where $f_0$ and $\alpha$ are real parameters~\cite{Jhingan:2008ym,Odintsov0708,Koivisto,Nojiri:2010pw,Bamba1104,ZS,EPV,Bahamonde:2017sdo}.
  An explicit mechanism to screen the
cosmological constant in these models has been investigated in~\cite{Nojiri:2010pw,Bamba1104,ZS}. Its local formulation, including an additional Gauss-Bonnet term, has been explored in~\cite{Non-local-FR}.

Another type of non-local infrared modifications of general relativity that includes the $\Box^{-1}$ operator squared has been proposed in~\cite{Foffa:2013vma} to describe the dark energy dominated epoch.
   Cosmological perturbations and structure formation in these models have been studied in~\cite{Dirian:2014ara} (see also~\cite{Maggiore:2016gpx}).  It has been shown that they allow to describe the CMB, BAO, and supernova data and fit these datasets at the same confidence level as $\Lambda$CDM~\cite{Dirian:2014bma,Belgacem:2017cqo}.
In the paper~\cite{Capozziello:2008gu},  non-local models with the Gauss-Bonnet term of a quite general form have been proposed, and  accelerating cosmological solutions have been studied\footnote{Note that there is another class of non-local models with Gauss-Bonnet term, which includes analytic functions of the d'Alembertian operator~\cite{Koshelev:2013lfm}.}. Also, a localization procedure that transforms a non-local model with the inverse d'Alambert operator acting on the Gauss-Bonnet term into a model of string-inspired scalar-Gauss-Bonnet gravity has been proposed in~\cite{Capozziello:2008gu}. In~\cite{Cognola:2009jx} this model has been modified by changing the $\Box^{-1}$ operator for a $(\Box-m^2)^{-1}$ operator, where $m$ is a new massive parameters. This modification allows to obtain de Sitter solutions with  constant scalar fields that are absent in the original model~\cite{Capozziello:2008gu}.

In the present paper, we will continue to investigate the class of non-local models proposed in~\cite{Capozziello:2008gu}, and check for the existence of de Sitter and power-law solutions. We will find the conditions on the parameters of the model that are both necessary and sufficient for the existence of de Sitter solutions. We will also look for power-law solutions and prove that models with de Sitter solutions have no power-law solutions, whereas models without de Sitter solutions can have power-law ones. This result is important in relation with the characterization of the different epochs of the Universe evolution, as described above. Both de Sitter and power-law solutions are found in analytic form.

\section{Non-local models with the Gauss-Bonnet term and their local formulation}
\subsection{Action}
Let us consider the following non-local model with  Gauss-Bonnet term $\mathcal{G}$:
\begin{equation}
\label{NLGBaction}
S_{NL}=\int dx^4\sqrt{-g}\left[\frac{M_{\mathrm{Pl}}^2}{16\pi}R+C\mathcal{G}^{n_1}\Box^{-n_2}\mathcal{G}^{n_3}-\Lambda\right],
\end{equation}
where $M_{\mathrm{Pl}}$ is the  Planck mass, $C$ and $\Lambda$ are constants, $n_k$  natural numbers, $\Box$ the d'Alembertian operator in the metric $g_{\mu\nu}$, with determinant $g$, and the Gauss-Bonnet term is given by
\begin{equation*}
\mathcal{G}=R^2-4R_{\mu\nu}R^{\mu\nu}+R_{\mu\nu\alpha\beta}R^{\mu\nu\alpha\beta}.
\end{equation*}
Using a localization procedure that is analogous to one proposed in~\cite{Odintsov0708,Capozziello:2008gu}, we introduce scalar fields and get
the corresponding local model
\begin{equation}
S_{L}=\int dx^4\sqrt{-g}\left[\frac{M_{\mathrm{Pl}}^2}{16\pi}R+C\mathcal{G}^{n_1}\phi+\xi\left(\Box^{n_2}\phi-\mathcal{G}^{n_3}\right)-\Lambda\right].
\end{equation}

When $n_2>1$, the scalars fields $\xi$ and $\phi$ can be expressed  in terms of $2n_2$ scalar fields $\xi_j$ and $\phi_j$  that satisfy second order equations. Indeed,
 \begin{equation*}
    S_{L}=\int dx^4\sqrt{-g}\left[\frac{M^2_{Pl}}{16\pi}R+C\mathcal{G}^{n_1}\phi_{n_2}+\xi_1(\Box\phi_1-\mathcal{G}^{n_3})+
    \sum_{j=2}^{n_2}\xi_j(\Box\phi_j-\phi_{j-1})-\Lambda\right]\!.
    \end{equation*}
    Varying this action under $\xi_j$, we get
   \begin{equation}
   \label{Equphi}
    \begin{split}
    \Box\phi_1=\mathcal{G}^{n_3}, & \qquad j=1, \\
 \Box\phi_j=\phi_{j-1}, &   \qquad j=2,\dots,n_2.
 \end{split}
   \end{equation}
    Also, this action can be presented in the form
        \begin{equation*}
  S_{L1}=\!\int\! dx^4\sqrt{-g}\left[\frac{M^2_{Pl}}{16\pi}R+C\mathcal{G}^{n_1}\phi_{n_2}-\xi_1\mathcal{G}^{n_3}+\sum_{j=1}^{n_2}\phi_j\Box\xi_j-
  \sum_{j=1}^{n_2-1}\xi_{j+1}\phi_{j}-\Lambda\right]\!,
  \end{equation*}
and variation under $\phi_{J}$  leads to
\begin{equation}\label{Equxi}
\begin{split}
\Box\xi_{n_2}={}-  C\mathcal{G}^{n_1}, &\qquad j=n_2, \\
 \Box\xi_{j}=\xi_{j+1},   &\qquad j=1,\dots,n_2-1.
\end{split}
\end{equation}

The action $S_{L}$ can be linearized with respect to the Gauss-Bonnet term, by adding one more scalar field to it (see~\cite{Cognola:2006eg}). Let us consider the part of action $S_L$ that includes the Gauss-Bonnet term, namely
\begin{equation}
    S_{fGB}=\int dx^4\sqrt{-g}\left[C\mathcal{G}^{n_1}\phi_{n_2}-\xi_1\mathcal{G}^{n_3}\right].
\end{equation}
To linearize this action with respect to $\mathcal{G}$
we introduce a scalar field $\sigma$ and
\begin{equation}
    f(\sigma)=C\sigma^{n_1}\phi_{n_2}-\xi_1\sigma^{n_3},
\end{equation}
and thus get the following equivalent action:
\begin{equation}
\begin{split}
    S_{GB\sigma}&=\int dx^4\sqrt{-g}\left[\frac{df}{d\sigma}(\mathcal{G}-\sigma)+f\right]={}\\
    =&\!\int\! dx^4\sqrt{-g}\left[\left(n_1C\sigma^{n_1-1}\phi_{n_2}-n_3\xi_1\sigma^{n_3-1}\right)(\mathcal{G}-\sigma)
    +C\sigma^{n_1}\phi_{n_2}-\xi_1\sigma^{n_3}\right]\!.
    \end{split}
\end{equation}
Varying over $\sigma$, one obtains that $\sigma=\mathcal{G}$ and the action $S_{fGB}$. Note that the scalar field $\sigma$ is not dynamical, because it has no kinetic energy term.

As a consequence, the initial action $S_{NL}$ can be written in the following scalar-tensor form:
\begin{equation}
S=\int dx^4\sqrt{-g}\left[\frac{M^2_{Pl}}{16\pi}R+F\mathcal{G}-V-\sum_{k=1}^{n_2}g^{\mu\nu}\partial_\mu{\xi}_k\partial_\nu\phi_k\right],
\end{equation}
where we use the following  re-designations
\begin{eqnarray}
  F &=&n_1{C}\sigma^{n_1-1}\phi_{n_2}-n_3{\xi}_1\sigma^{n_3-1}\,, \\
   V&=&{}-{C}\sigma^{n_1}\phi_{n_2}(1-n_1)-{\xi}_1\sigma^{n_3}(n_3-1)+\sum_{k=1}^{n_2-1}{\xi}_{k+1}\phi_k+
   \Lambda\,.
\end{eqnarray}
A similar action with one scalar field non-minimally coupled with the Gauss-Bonnet term has been studied in~\cite{Cognola:2006eg,Cognola:2006sp}.

Varying the  local action $S$ thus obtained, we get the following equations
\begin{equation}
\label{EinstEqs}
\begin{split}
 & \left[\frac{M^2_{Pl}}{16\pi}-4\Box F\right]\left[R^{\mu\nu}-\frac12g^{\mu\nu}R\right]
+\frac{g^{\mu\nu}}{2}\left(\sum_{k=1}^{n_2}\partial_\sigma\phi_k\partial^\sigma{\xi}_k\right)
-\sum_{k=1}^{n_2}\partial^\mu\phi_k\partial^\nu\xi_k=\\
    &=F\left[\frac12g^{\mu\nu}\mathcal{G}-2R R^{\mu\nu}+4 R^\mu_\rho R^{\nu\rho}-2 R^{\mu\rho\sigma\tau}R^\nu_{\rho\sigma\tau}+4R^{\mu\rho\sigma\nu}R_{\rho\sigma}\right]-\frac{g^{\mu\nu}}{2}V+{}\\
    &2R(D^\mu D^\nu F)-4(D_\rho D^\mu F) R^{\nu\rho}-4(D_\rho D^\nu F)R^{\mu\rho}+{}\\
    &4g^{\mu\nu}(D_\rho D_\sigma F)R^{\rho\sigma}-4(D_\rho D_\sigma F) R^{\mu\rho\nu\sigma}.
\end{split}
\end{equation}
Variation under $\sigma$ leads to
\begin{equation}
\left[n_1(n_1-1){C}\sigma^{n_1-2}\phi_{n_2}+n_3(n_3-1){\xi}_1\sigma^{n_3-2}\right]\left(\mathcal{G}-\sigma\right)=0,
\quad \Rightarrow \quad \mathcal{G}=\sigma\,.
\end{equation}

Let us consider the trace of Eq.~\eqref{EinstEqs}.
After setting $R^{\mu\rho\sigma\nu}=R^{\nu\rho\mu\sigma}$ and $g_{\mu\nu}R^{\nu\rho\mu\sigma}R_{\rho\sigma}=R_\mu^{\rho\mu\sigma}R_{\rho\sigma}=R^{\rho\sigma}R_{\rho\sigma}$,
we get
\begin{equation}
g_{\mu\nu}F\left[\frac12g^{\mu\nu}\mathcal{G}-2R R^{\mu\nu}+4 R^\mu_\rho R^{\nu\rho}-2 R^{\mu\rho\sigma\tau}R^\nu_{\rho\sigma\tau}+4R^{\mu\rho\sigma\nu}R_{\rho\sigma}\right]=0
\end{equation}

Using $R^{\mu\nu}D_\mu D_\nu F=R\Box F$ and
\begin{equation*}
{}-8(D_\rho D^\nu F)R^{\rho}_\nu+16(D_\rho D_\sigma F)R^{\rho\sigma}-4(D_\rho D_\sigma F) R_\nu^{\rho\nu\sigma}=4(D_\rho D_\sigma F) R^{\rho\sigma},
\end{equation*}
we obtain the trace equation:
\begin{equation}\label{TraceEq}
\frac{M^2_{Pl}}{16\pi}R-\left(\sum_{k=1}^{n_2}\partial_\sigma\phi_k\partial^\sigma{\xi}_k\right)-2V
-R(\Box F)=0\,.
\end{equation}

\section{Cosmological equations}
\subsection{The FLRW metric}
Let us consider the spatially flat FLRW universe with metric
\begin{equation}
\label{metric}
ds^2={}-dt^2+a^2(t)\left(dx_1^2+dx_2^2+dx_3^2\right),
\end{equation}
From this metric, one gets ($i,j,m,l=1,2,3$):
\begin{equation*}
R^{i0j0}=R^{0i0j}=-R^{0ij0}=-R^{i00j}={}-\frac{\left[\dot{H}+H^2\right]}{a^2}\delta_{ij}, \,   R^{ijml}=\frac{H^2}{a^4}(\delta_{im}\delta_{lj}-\delta_{il}\delta_{mj}),
\end{equation*}
\begin{equation*}
\Gamma^0_{ij}=a^2H\delta_{ij},\qquad \Gamma^i_{0j}=\Gamma^i_{j0}=H\delta^i_{j},
\end{equation*}
\begin{equation*}
\begin{split}
&R^{00}={}-3\left(\dot{H}+H^2\right),\quad R^{ij}=\frac{\left[\dot{H}+3H^2\right]}{a^2}\delta_{ij}, \\ & R=6(\dot H+2H^2),\quad {\cal G}=24H^2\left(\dot H +H^2\right),
\end{split}
\end{equation*}
where the Hubble parameter is $H=\dot{a}/a$, and the dots mean time derivatives.

We will assume that all scalar fields depend on time only, and thus get the following expressions, including components equations
\begin{equation}
\Box F={}-3H\dot F-\ddot F,
\end{equation}
\begin{equation}
D_\rho D_\sigma F=\partial^2_{\rho\sigma}F-\Gamma^j_{\rho\sigma}\partial_j F,
\end{equation}
from where
\begin{equation}
D^i D^i F={}-\frac{H}{a^2}\dot F,\quad D_iD_i F={}-a^2H\dot{F},
\end{equation}
\begin{equation}
D^0D^0 F=D_0D_0F=\ddot{F},
\qquad D^iD^0F=D^0D^iF=0\,.
\end{equation}

\subsection{Field and Friedmann equations}

The field equations (\ref{Equphi}) and (\ref{Equxi}) in the FLRW metric~(\ref{metric}) have the following form
\begin{equation}\label{Equfields}
    \ddot\phi_k=-3H\dot\phi_k+{\cal G}\frac{\partial F}{\partial \xi_k}-\frac{\partial V}{\partial \xi_k},\qquad
    \ddot\xi_k=-3H\dot\xi_k+{\cal G}\frac{\partial F}{\partial \phi_k}-\frac{\partial V}{\partial \phi_k}\,.
\end{equation}
Eqs.~(\ref{EinstEqs}) in the FLRW metric read as follows
\begin{equation}
3H^2\frac{M^2_{Pl}}{16\pi}-\frac{1}{2}\sum_{k=1}^{n_2}\left(\dot{\phi}_k\dot{\xi}_k\right)-\frac12 V=-12H^3\dot{F},
\label{00}
\end{equation}
\begin{equation}
-\left( 3H^2+2\dot{H} \right)\frac{M^2_{Pl}}{16\pi}-8H \left( {H}^{2}+\dot{H}\right) {\dot{F}}-4{H}^{2}{\ddot{F}}-\frac{1}{2}\sum_{k=1}^{n_2}\dot{\phi}_k\dot{\xi}_k+\frac{V}{2}=0.\label{11}
\end{equation}
 Subtracting \eqref{11} from \eqref{00}, we get
 \begin{equation}
 8{H}^{3}{\dot{F}}-4{\Box{F}}{H}^{2}+\frac{3M^2_{Pl}}{8\pi}{H}^{2}+8H\dot{F}\dot{H}+\frac{M^2_{Pl}}{8\pi}\dot{H}-V=0.\label{int}
\end{equation}
Note that Eqs.~(\ref{11}) and (\ref{int}) are third order differential equations with respect to the Hubble parameter.

Let us show that Eq.~(\ref{11}) is a consequence of Eqs.~(\ref{Equfields}) and (\ref{00}).
Differentiating Eq.~(\ref{00}) with respect to $t$, we obtain
\begin{equation}
\label{d00}
 6H\dot H\frac{M^2_{Pl}}{16\pi}-\frac{1}{2}\left[\sum_{k=1}^{n_2}\left(\ddot{\phi}_k\dot{\xi}_k+\dot{\phi}_k\ddot{\xi}_k\right)+ \dot V\right]=-36H^2\dot{H}\dot{F}-12H^3\ddot{F}.
\end{equation}
Taking into account Eqs.~(\ref{Equfields}) we modify Eq.~(\ref{d00}), using
\begin{equation}
 \sum_{k=1}^{n_2}\left(\ddot{\phi}_k\dot{\xi}_k+\dot{\phi}_k\ddot{\xi}_k\right)+ \dot V={}-6H\sum_{k=1}^{n_2}\left(\dot{\phi}_k\dot{\xi}_k\right)+{\cal G}\dot{F}.
\end{equation}
Dividing Eq.~(\ref{d00}) by $3H$ and adding it up with  Eq.~(\ref{00}), we get Eq.~(\ref{11}). So, to obtain solutions of the  models we are here considering, it is enough to solve the corresponding field equations together with Eq.~(\ref{00}).

\section{Search for de Sitter solutions}

If the Hubble parameter is a constant: $H=H_0$, then the Gauss-Bonnet term reads $\mathcal{G}=24H_0^4\equiv \mathcal{G}_0$ and $\sigma=\mathcal{G}_0$.
As a consequence, the corresponding field equations (\ref{Equphi}) get transformed into the following system of linear first order differential equations, with constant coefficients,
\begin{equation}
   \label{EquphidS}
    \begin{split}
    \dot\phi_1&=\psi_1,\\
    \dot{\psi}_1&={}-3H_0\psi_1-\mathcal{G}_0^{n_3}, \\
    \dot{\phi}_j&=\psi_{j},    \qquad j=2,\dots,n_2,\\
 \dot{\psi}_j&={}-3H_0\psi_j-\phi_{j-1},    \qquad j=2,\dots,n_2.
 \end{split}
 \end{equation}
The system (\ref{EquphidS}) has the following solution
\begin{equation}
    \phi_j=P_j(t)e^{-3H_0t}-\frac{\mathcal{G}_0^{n_3}}{j!(3H_0)^j}t^{j}+\tilde{P}_j(t),
\end{equation}
where $P_j(t)$ and $\tilde{P}_j(t)$ are $(j-1)$-degree polynomials of $t$ with coefficients that include $2j$ arbitrary parameters.

Analogously, the system (\ref{Equxi}) acquires the following form
\begin{equation}
\begin{split}
\ddot{\xi}_{j}+3H_0\dot{\xi}_{j}+\xi_{j+1}=0,   &\qquad j=1,\dots,n_2-1, \\
\ddot{\xi}_{n_2}+3H_0\dot{\xi}_{n_2}- C\mathcal{G}_0^{n_1}=0, &
\end{split}
\end{equation}
and the solution reads
\begin{equation}
    \xi_j=Q_j(t)e^{-3H_0t}+\frac{C\mathcal{G}_0^{n_1}}{(n_2-j+1)!(3H_0)^{n_2-j+1}}t^{(n_2-j+1)}+\tilde{Q}_j(t),
\end{equation}
where $Q_j(t)$ and $\tilde{Q}_j(t)$ are  polynomials in $t$ of degree $(n_2-j)+1$.

To check for the existence of de Sitter solutions, one must substitute the  solutions of the field equations thus obtained into Eqs.~(\ref{00}) and (\ref{int}). In the case when $n_1=n_2=n_3=1$ the Sitter solutions have been found in~\cite{Capozziello:2008gu}.
In this paper, we consider the case $n_2=2$.

The field equations
\begin{equation}
 -\ddot{\phi}_1-3H_0\dot{\phi}_1=\mathcal{G}_0^{n_3},\qquad -\ddot{\phi}_2-3H_0\dot{\phi}_2=\phi_1
\end{equation}
and
\begin{equation}
 -\ddot{\xi}_2-3H_0\dot{\xi}_2=-C\mathcal{G}_0^{n_1},\qquad -\ddot{\xi}_1-3H_0\dot{\xi}_1=\xi_2
\end{equation}
 have the following solutions:
\begin{equation}
\phi_1=A_1e^{-3H_0t}-\frac{\mathcal{G}_0^{n_3}}{3H_0}t+B_1\,,
\end{equation}
\begin{equation}
\phi_2=\left(\frac{A_1}{3H_0}t+A_2\right)e^{-3H_0t}+\frac{\mathcal{G}_0^{n_3}}{18H_0^2}t^2
-\left(\frac{\mathcal{G}_0^{n_3}}{27H_0^3}+\frac{B_1}{3H_0}\right)t+B_2\,,
\end{equation}
\begin{equation}
\xi_1=\left(\frac{C_1}{3H_0}t+C_2\right)e^{-3H_0t}-\frac{C\mathcal{G}_0^{n_1}}{18H_0^2}t^2
+C\left(\frac{\mathcal{G}_0^{n_1}}{27H_0^3}-\frac{D_1}{3H_0}\right)t+D_2\,,
\end{equation}
\begin{equation}
\xi_2=C_1e^{-3H_0t}+C\frac{\mathcal{G}_0^{n_1}}{3H_0}t+CD_1\,,
\end{equation}
where $A_i$, $B_i$, $C_i$, and $D_i$ are integration constants.

Substituting these expressions into Eq.~(\ref{00}),
 \begin{equation}\label{dS00}
 3H_0^2\frac{M^2_{Pl}}{16\pi}-\frac{1}{2}\left(\sum_{k=1}^{n_2}\dot{\phi}_k\dot{\xi}_k\right)-\frac12 V=-12H_0^3\dot{F},
\end{equation}
we see that this equation can be satisfied only if $n_1+n_3=4$. Also, we get the following restriction to the integration constants
\begin{equation*}
A_1=0, \quad B_1 ={}- \frac{331776(n_1-2)H_0^{16-8n_1}D_1+24^{n_1}442368H_0^{14-4n_1}}{(n_1-2)},
\end{equation*}
\begin{equation*}
C_1=0, \quad C_2 ={}-\frac{24^{2n_1}(2n_1-1)A_2CH_0^{8(n_1-2)}}{331776(2n_1-7)}.
\end{equation*}
These restrictions are not valid for $n_1=2$.

Also, we have the additional to connect the values of the parameters of the solutions, with~$\Lambda$
\begin{equation}
\begin{split}
\Lambda&={}-\frac{3H_0^2M_{Pl}^2}{8\pi}-\frac{8192C(13n_1+4)H_0^{12}}{(n_1-2)}\\
&{}+24^{-n_1}331776D_2(n_1-3)H_0^{16-4n_1}+24^{n_1}CB_2(n_1-1)H_0^{4n_1}\\
&{}-\frac{24^{-n_1}73728(5n_1-4)CD_1H_0^{14-4n_1}}{n_1-2}
-24^{-2n_1}331776CD_1^2H_0^{16-8n_1}.
\end{split}
\end{equation}
Consequently, the value of $\Lambda$ fixes the value of one of the integration constants: $B_2$ for $n_1=3$ or $D_2$ for $n_1=1$.

Summing up, we do get explicitly de Sitter solutions for models with  $n_1=1$, $n_2=2$, $n_3=3$ and $n_1=3$, $n_2=2$, $n_3=1$. And we have also discovered that models with $n_2=2$ and other values of $n_1$ and $n_3$ do not have de Sitter solutions\footnote{Straightforward substitution of the field expressions when $n_1=n_3=2$ already proves the absence of the de Sitter solutions in this case.}. In the next section we will check the existence of power-low solutions for those models.

\section{Power Law solutions}

The search of power-law solutions with $H=J/t$ is more complicated.
We consider the case when $n_1$ or $n_3$ is equal to $1$.
If $n_1=1$ and $n_3=1$, then
 \begin{equation}
V=\xi_2\phi_1+\Lambda,\qquad F=C\phi_2-\xi_1,
\label{expressions}
\end{equation}
with the following form for the field equations
\begin{equation}
\Box\phi_1=\mathcal{G},\qquad \Box\phi_2=\phi_1,
\label{BoxPhi}
\end{equation}
\begin{equation}
\Box\xi_2=-C\mathcal{G},\qquad \Box\xi_1=\xi_2,
\label{BoxXi}
\end{equation}
where $\mathcal{G}=24(J-1)J^3/t^4$.

Using these formulas, we immediately obtain the form of Eq.~\eqref{int}
\begin{equation}\label{For Check}
  -2\left(3{H}^{2}+{\dot{H}} \right) \frac{M^2_{Pl}}{16\pi}-8H\left({H}^{2}+{\dot{H}} \right) (C\dot{\phi}_2-\dot{\xi}_1)+4(C{\phi}_1-{\xi}_2){H}^{2}+\xi_2\phi_1=0\,.
  \end{equation}
The model with $n_1=1$ and $n_3=1$ yields power law solutions with $H=J/t$ at $J=2/3$ and $J=3$. These solutions exist at $\Lambda=0$ only.
The corresponding scalar fields admit two types of expressions. The first type of solutions corresponds to
\begin{eqnarray}
\nonumber \phi_1&=&-{\frac {{C_1}\,{t}^{-3J+1}}{3J-1}}+4\,{\frac {{{J}}^{3}}{{t}^{2}}}
-\frac{M^2_{Pl}}{32\pi}\frac {\left( 3J+1 \right)}{J\,C \left( J-1 \right) },\\
\nonumber \phi_2&=&\frac{M^2_{Pl}}{64\pi}\,{\frac {{t}^{2}}{{J}\,C \left( {J}-1 \right) }}
-{\frac {{C_3}\,{t}^{-3J+1}}{C \left( 3J-1\right) }}
-\frac{{C_1}\,{t}^{3-3J}}{6(3\,J^2-4\,{J}+1)}
-4\,{\frac{{{J}}^{3}\ln\left( t \right)}{3J-1}}+{C_4},\\
\nonumber \xi_1&=&4\,{\frac {C{{J}}^{3} }{3J-1}}\ln  \left( t \right)
-{\frac {{C_3}\,{t}^{-3J+1}}{3J-1}}+{C_2},\\
\nonumber \xi_2&=&-4\,{\frac {C{{J}}^{3}}{{t}^{2}}},
\end{eqnarray}
where in the case $J=2/3$, $C_1={\frac {7168}{729\,{C_3}}}$,  while in the case $J=3$, either $C_1=0$  or $C_3=0$.

Another type of solutions, with the same Hubble parameter, is given by
\begin{eqnarray}
\nonumber \phi_1&=&4\,{\frac {{{J}}^{3}}{{t}^{2}}},\\
\nonumber \phi_2&=&-{\frac {{C_3}\,{t}^{-3J+1}}{3J-1}}-\,{\frac{4J^3 }{3J-1}}\ln\left( t \right)+{C_4},\\
\nonumber \xi_1&=&4\,{\frac {C{{J}}^{3}\ln\left( t \right) }{3J-1}}-{
\frac {C{C_3}\,{t}^{-3J+1}}{3J-1}}-\frac{M^2_{Pl}}{64\pi}\,{\frac
{{t}^{2}}{{J}\left( {J}-1 \right) }}-{\frac {{t}^{3
-3J}{C_1}}{6(3\,J^2-4\,{J}+1)}}+{C_2},\\
\nonumber \xi_2&=&-{\frac {{C_1}\,{t}^{-3J+1}}{3J-1}}-4\,{\frac {
C{{J}}^{3}}{{t}^{2}}}+\frac{M^2_{Pl}}{32\pi}\,{\frac {\left( 3J+1
 \right) }{{J}\, \left( {J}-1 \right) }},
\end{eqnarray}
where in the case $J=2/3$ we have the additional condition $C_1={}-{\frac {7168\,C}{729\,{C_3}}}$, while
in the case $J=3$, either $C_3=0$ or $C_1=0$. Other $C_i$ are arbitrary constants.
Note that the form of the solutions obtained excludes a few values of $J$, which must be checked separately.
There is no solution with $J=1/3$.
In the case $J=1$ that corresponds to ${\cal G}=0$ we do not seek solutions.

Let us consider the case $n_1=1$ and $n_3>1$. The expressions for $\phi_2(t)$ do not include a logarithmic term, whereas $\xi_1(t)$ has a logarithmic term, which cannot be remove by the choice of integration constants. Substituting the field expressions thus obtained into Eq.~(\ref{00}), we get that the logarithmic term of the equation has the following coefficient
\begin{equation*}
\frac{24^{n_3}(2(n_3-1))CJ^3(J-1)^{n_3}J^{3n_3}(4n_3+J-1)}{(J-1)(3J-1)}t^{-4n_3}.
\end{equation*}
This coefficient is equal to zero only for $J ={}-4n_3+1<0$. Such case is not interesting because the corresponding Hubble parameter is strictly negative. As a consequence, the model with $n_1=1$ and $n_3>1$ admits power-law solutions only for  values of $J$ such that a denominator of some field expression is zero. At the same time, there is no solution with $J=1/3$ for models with $n_1=1$ and $n_3>1$.

\section{Conclusion}

In this paper we have analyzed a set of non-local cosmological models with the Gauss-Bonnet term given by the action (\ref{NLGBaction}). We have shown how this action can be transformed into a local one and, after that, we have looked for de Sitter and power-law solutions for the corresponding models in the specific case $n_2=2$.

The result we have obtained is that de Sitter solutions exist only in these two cases: for $n_1=1$ and $n_3=3$, or for $n_1=3$ and $n_3=1$.
We have seen that these models yield no power-law solutions. Moreover, we have shown that, if $n_1=1$ and $n_3>1$ (or $n_1>1$ and $n_3=1$, respectively), then power-law solutions do not exist. In the case $n_1=n_3=1$, power-law solutions with $H=J/t$ exist only for $J=2/3$ and $J=3$. Therefore, the model with $n_1=1$, $n_2=2$, and $n_3=1$, without additional matter, is suitable in order to describe the matter-dominated phase of the Universe evolution that corresponds to $J=2/3$.

As an extension of this work, we plan to consider the stability of the  de Sitter solutions obtained and the possibility to describe dark energy in  models of this type.  Also, it is naturally assumed that non-local modifications of gravity can be important at the very high energy scales that correspond to inflation. In this respect, note that the models here with de Sitter solutions may serve for the construction of
eternal inflation. It is thus an interesting and non-trivial problem to try to generalize the known inflationary scenarios by adding a non-local gravity term~\cite{Odintsov0708,Koshelev:2017tvv}. We plan to consider the possibility to use the here considered models of non-local Gauss-Bonnet gravity for the generalization of local inflationary scenarios.

\medskip

\section*{Acknowledgements}

This work has been partially supported by  MINECO (Spain), Project  FIS2016-76363-P, and by the CPAN Consolider Ingenio 2010 Project.
Research of E.P. and S.V. is supported in
part by the RFBR grant 18-52-45016.

\end{document}